\documentclass[twocolumn,aps,pra,floatfix]{revtex4-2}

\usepackage{amsmath}
\usepackage{txfonts}
\usepackage{microtype}
\usepackage{graphicx}
\usepackage{color}
\usepackage{ulem}

\begin{document}

\title{Sub-cycle time-resolved nondipole dynamics in tunneling ionization}

\author{Michael Klaiber}\email{klaiber@mpi-hd.mpg.de}
\affiliation{Max-Planck-Institut f\"ur Kernphysik, Saupfercheckweg 1, 69117 Heidelberg, Germany}
\author{Karen Z. Hatsagortsyan}\email{k.hatsagortsyan@mpi-hd.mpg.de}
\affiliation{Max-Planck-Institut f\"ur Kernphysik, Saupfercheckweg 1, 69117 Heidelberg, Germany}
\author{Christoph H. Keitel}
\affiliation{Max-Planck-Institut f\"ur Kernphysik, Saupfercheckweg 1, 69117 Heidelberg, Germany}

\date{\today}

\begin{abstract}

The electron nondipole dynamics in tunneling ionization in an elliptically polarized laser field is investigated theoretically using a relativistic Coulomb-corrected strong field approximation (SFA) based on the eikonal approximation of the Klein-Gordon equation. We calculate   attoclock angle-resolved light-front momentum distributions at different ellipticities of the laser field in quasistatic and nonadiabatic regimes and analyze them with an improved Simpleman model. The nondipole correlations between longitudinal and transverse momentum components are examined. Deviations of the photoelectron momentum distribution calculated via SFA with respect to  the available experimental results as well as with the improved Simpleman model are  discussed and interpreted  in terms of  nonadiabatic as well as Coulomb effects in the continuum and under-the-barrier.  The favorable prospects of an experimental observation are discussed.

\end{abstract}

\maketitle

\section{Introduction}

High precision measurements in strong-field atomic physics allow to detect nondipole features in photoelectron momentum distribution (PMD) at laser intensities far below the relativistic regime \cite{Smeenk_2011,Ludwig_2014,Maurer_2018,Willenberg_2019,Hartung_2019,Haram_2019,Grundmann_2020,Hartung_2021,Lin_2021,Lin_2021a}.
The leading nondipole effect in tunneling ionization is due to the laser magnetic field and  results in imparting the photoelectron a momentum along the laser propagation direction, which has consequences for the partitioning of the absorbed photon momentum between the photoelectron and the parent ion \cite{Smeenk_2011,Klaiber_2013,Chelkowski_2014,Cricchio_2015,Bandrauk_2015,Ivanov_2015,Ivanov_2016,Simonsen_2016,Chelkowski_2017,
Keil_2017,Tao_2017,He_2017,He_2021}. The electron energy resolution of state-of-the-art detection techniques \cite{COLTRIM} is of about meV, which corresponds to a momentum resolution of about 0.01~a.u. This means that a nondipole shift of a longitudinal momentum $p_k\sim ca_0^2$ \cite{RMP_2012} can be detected in a laser field with $a_0\sim 10^{-2}$ corresponding to a laser intensity $I\sim 10^{14}$ W/cm$^2$ at 800 nm wavelength. Here, $a_0=E_0/(c\omega)$ is the relativistic laser field parameter \cite{Ritus_1985}, with the laser field amplitude $E_0$, frequency $\omega$, and the speed of light $c$. Atomic units are used throughout. With the recent advancement of the strong field laser technique  into the mid-IR region up to wavelengths of the order of 10 $\mu$m \cite{Wolter_2015x}, the nondipole effects become measurable at even lower laser intensities. The Lorentz force effect matters not only in the continuum but also during the sub-barrier tunneling dynamics, inducing an additional longitudinal momentum shift  $I_p/(3c)$  \cite{Klaiber_2013,Hartung_2019}, with the ionization potential $I_p$. The latter is increased by sub-barrier Coulomb corrections \cite{He_2021}.

In a linearly polarized laser field,  the drift of the electron induced by the laser magnetic field is known to suppress the recollision and related phenomena, see e.g. \cite{Dammasch_2001,Kohler_adv,Klaiber_2017}. At restrained recollisions, the interplay between the Coulomb, ellipticity, and nondipole effects in the continuum induces specific structures in PMD \cite{Forre_2006,Liu_2012,Ludwig_2014,Maurer_2018,Danek_2018b,Danek_2018,Danek-II_2018,Danek_2019,Willenberg_2019,Maurer_2021}.
In an elliptically polarized laser field close to circular  the Coulomb field of the atomic core disturbs the photoelectron motion in the continuum mostly near the tunnel exit, however, it also modifies the sub-barrier dynamics \cite{He_2021}.

While in first experiments \cite{Smeenk_2011,Hartung_2019,Haram_2019} the average of the longitudinal momentum shift was in the attention of investigation, the recent experiment of Ref.~\cite{Willenberg_2019} provides a sub-cycle time-resolved study, and the experiment of Ref.~\cite{Hartung_2021} investigates nondipole correlations between longitudinal-transverse momentum components in the ionized wave packet. The nondipole effects have been observed also in high-order above-threshold ionization \cite{Lin_2021}, and the photoelectron energy peaks shift against the radiation
pressure has been shown in the experiment  \cite{Lin_2021a}. The results of these experiments have raised significant interest of theory, addressing different aspects of the nondipole phenomena, in particular investigating the nonadiabatic \cite{Ni_2020}, and Coulomb effects \cite{Haram_2020,Ma_2021}, as well as  the inter-cycle interference structure in the PMD in the nondipole regime \cite{Lund_2021,Brennecke_2021}.

In this paper we investigate theoretically the electron nondipole dynamics in an elliptically polarized laser field in detail. A relativistic  strong field approximation (SFA) is employed and Coulomb corrections in the continuum as well as during tunneling are included in the eikonal approximation. Main attention is devoted to the attoclock angle-resolved light-front  momentum distributions and on possible different correlations between longitudinal-transverse momentum components. The role of nonadiabatic and Coulomb effects in the continuum and during tunneling as well as their interplay are analyzed. The Simpleman model \cite{Corkum_1993} is improved, including nonadiabatic and Coulomb corrections, for an intuitive interpretation of the PMD features within the SFA theory. The light-front momentum as a choice for an observable is underlined as being especially suitable for exploring the deviations from the Simpleman model due to nonadiabatic and Coulomb corrections.

\section{Theoretical approach}\label{Sec:parameters}

We employ a relativistic Coulomb-corrected SFA (CCSFA) based on  the Klein-Gordon equation, where the Coulomb potential of the atomic core is accounted for using the eikonal approximation \cite{Klaiber_2013_I,Klaiber_2013_II}. The  ionization amplitude is calculated in the dressed partition~\cite{Klaiber_2006gauge}, neglecting small spin effects \cite{Klaiber_2005}:
\begin{eqnarray}
m_\mathbf{p}=-i\int dt \langle\psi_\mathbf{p}(t)|H_i|\phi(t)\rangle
\label{amplitude}
\end{eqnarray}
with the interaction Hamiltonian $H_i=\mathbf{r}\cdot\mathbf{E}(\eta)$. The laser field is elliptically polarized:
\begin{equation}
\label{field}
   \mathbf{E}=-\frac{E_0}{\sqrt{1+\epsilon^2}} \left[\mathbf{e}_x\cos(\omega\eta)+\epsilon \mathbf{e}_y\sin(\omega\eta)]\right],
\end{equation}
where $\epsilon$ is the ellipticity, $\eta=t-\hat{\mathbf{k}}\cdot \mathbf{r}/c=t-z/c$, $\hat{\mathbf{k}}$ the unit vector along the laser propagation direction, $\phi(\mathbf{r},t)=c_a\phi_0(\mathbf{r},t)\phi_1(\mathbf{r},t)$ is the initial bound state of the electron in a Coulomb-potential $V=-Z/r$,  with charge $Z$ and an asymptotic expression at $r\gg 1/\kappa$:
\begin{eqnarray}
\label{bswf}
  \phi_0(\mathbf{r},t)&=& \frac{\exp(-\kappa r+i\kappa^2/2t)}{r}, \nonumber\\
  \phi_1(\mathbf{r})&=&\left(\sqrt{2}\kappa r\right)^{Z/\kappa},
\end{eqnarray}
with $c_a \equiv  \sqrt{\kappa/(2\pi)}$, and $\kappa=\sqrt{2I_p}$. We use the nonrelativistic bound state because the relativistic corrections scale as $I_p/c^2$ and are negligible for the applied conditions. The electron final state in the continuum is assumed to be the Coulomb-Volkov state in the eikonal approximation \cite{Klaiber_2013_II}:
\begin{eqnarray}
\psi_\mathbf{p}(\mathbf{r},t)=\frac{1}{(2\pi)^{3/2}}\exp[iS_0(\mathbf{r},t)+i S_1(\mathbf{r},t)].
\end{eqnarray}
 The applied eikonal approximation  is valid if the momentum change of the electron due to the Coulomb field is smaller with respect to the electron momentum via laser field. This is the case when hard recollisions do not play role, which exactly corresponds to the electron dynamics discussed in this paper, namely, ionization in an elliptically polarized laser field with $\epsilon \gtrsim 0.3$ \cite{Liu_2012,Maurer_2018}.
Here,
\begin{eqnarray}
S_0(\mathbf{r},t)&=&\left(\mathbf{p}+\mathbf{A}(\eta)- (\varepsilon/c-c)\hat{\mathbf{k}}\right)\cdot\mathbf{r}
+\int^{\infty}_\eta ds[\varepsilon(s)-c^2]\nonumber
\end{eqnarray}
is the Volkov-action, $\varepsilon=\sqrt{c^4+c^2\mathbf{p}^2}$ the electron energy,
\begin{equation}
\label{v-potential}
   \mathbf{A}(\eta)=\frac{E_0/\omega}{\sqrt{1+\epsilon^2}} \left[\mathbf{e}_x\sin(\omega\eta)-\epsilon \mathbf{e}_y\cos(\omega\eta)]\right]
\end{equation}
is the laser vector potential, and
\begin{eqnarray}
\varepsilon(s)=\varepsilon+\frac{\mathbf{p}\cdot\mathbf{A}(s)+\mathbf{A}(s)^2/2}{\Lambda}
\end{eqnarray}
is the electron energy in the laser field, with the integral of motion $\Lambda=\varepsilon/c^2-p_k/c$ and $p_k=\hat{\mathbf{k}}\cdot \mathbf{p}$.
Further, the Coulomb correction (CC) to the eikonal is
\begin{eqnarray}
S_1(\mathbf{r},t)=\int^\infty_{\eta}ds \frac{\varepsilon(s)}{\Lambda c^2}V(\mathbf{r}(s,\eta)),
\end{eqnarray}
with the electron relativistic trajectory
\begin{eqnarray}
\mathbf{r}(\eta',\eta)=\mathbf{r}+\frac{1}{\Lambda} \int^{\eta'}_\eta ds\left( \mathbf{p}+\mathbf{A}(s)+\hat{\mathbf{k}}\frac{\mathbf{p}\cdot\mathbf{A}(s)+\mathbf{A}(s)^2/2}{c\Lambda}\right). \nonumber\\
\end{eqnarray}

The ionization amplitude of Eq.~(\ref{amplitude}) consists of a 4-dimensional integral. After a coordinate transformation from $t$ to $\eta$, we solve it with the saddle-point method. Therefore, the integrand is exponentiated in cylindrical coordinates $\mathbf{r}=(\rho,\varphi,z)$
\begin{eqnarray}
m_\mathbf{p}=-i\int d\eta d\rho d\phi dz\exp(\zeta_0+\zeta_1),
\end{eqnarray}
where $\zeta_0=\ln(\rho c_a H_i\phi_0)-i S_0$ and $\zeta_1=\ln(\phi_1)-i S_1$. Consequently, we obtain the saddle-point equations:
\begin{eqnarray}
\partial_\eta \zeta_0(\rho,\varphi,z,\eta)&=&0\nonumber\\
\partial_\rho \zeta_0(\rho,\varphi,z,\eta)&=&0\nonumber\\
\partial_\varphi \zeta_0(\rho,\varphi,z,\eta)&=&0\nonumber\\
\partial_z \zeta_0(\rho,\varphi,z,\eta)&=&0.
\end{eqnarray}
In the saddle-point equations it was assumed that the first order term $\zeta_1$ is slowly varying and therefore neglected with respect to the $\zeta_0$-contribution. For a given final momentum $\mathbf{p}$ the saddle point equations are  solved numerically, obtaining the ionization amplitude:
\begin{eqnarray}
m_\mathbf{p}=-i\sqrt{\frac{(-2\pi)^4}{\det\partial_i\partial_j\zeta_{0,s}}}\exp[\zeta_{0,s}+\zeta_{1,s}],
\end{eqnarray}
where indices $i$ and $j$ run over the cylindrical coordinates and $\eta$. The corresponding momentum distribution is then calculated via
\begin{equation}\label{WWW}
  \frac{dw(\mathbf{p})}{d^3\mathbf{p}}=|m(\mathbf{p})|^2.
\end{equation}

\section{Simpleman Model}

In this section we extend the well-known Simpleman model \cite{Corkum_1993} into the relativistic domain   for spinless particle and further improve it in order to include the nondipole sub-barrier correction to the longitudinal momentum at the tunnel exit and its CC, the nonadiabatic corrections to the initial electron momentum at the tunnel exit, as well as Coulomb corrections due to the continuum motion in the quasistatic and in nonadiabatic regimes.

\subsection{Quasistatic regime}

In the Simpleman model we find the most probable trajectory for the ionized electron, and accordingly the most probable asymptotic momentum corresponding to the peak of PMD. In the quasistatic regime the ionized electron appears in the continuum at the tunnel exit (at the laser phase $\phi_i=\omega\eta_i$) with a vanishing momentum  $\textbf{p}_{\bot i}=0$, $p_{k i}=0$. Here $\textbf{p}_\bot$  is the transverse momentum component in the polarization plane, and $p_k$ the longitudinal component. Further the electron moves in the laser and Coulomb fields of the atomic core.

Firstly, we find the electron trajectory in a plane-wave laser field $\textbf{A}=\textbf{A}(\phi)$. In this field there are two integrals of motion following from the field symmetry, namely, on the field dependence only in the single variable $\phi$:
\begin{eqnarray}
\textbf{p}_\bot-\textbf{A}(\phi)&=&\textbf{p}_{\bot i}-\textbf{A}(\phi_i), \label{3}\\
\varepsilon - cp_k&=&\varepsilon_i - cp_{k i}= c^2\Lambda,
 \end{eqnarray}
with the initial energy $\varepsilon_i$  at $\phi=\phi_i$, $\phi=\omega\eta$. From the latter the final photoelectron momentum is derived (see e.g. Eq.~(A.10) in \cite{Maurer_2018}):
 \begin{eqnarray}
\textbf{p}_{\bot }&=&\textbf{p}_{\bot i}-\textbf{A}(\phi_i), \label{pxpx}\\
p_k&=&p_{k i}+\frac{\textbf{p}_{\bot }^2-\textbf{p}_{\bot i}^2}{2c\Lambda},\label{pzpz}
\end{eqnarray}
with $\Lambda\approx 1-p_{ki}/c\approx 1$. The latter can be expressed either via the initial transverse momentum $\textbf{p}_{\bot i}$ or via the asymptotic one $\textbf{p}_{\bot }$, which in the leading order of ${\cal O} (1/c)$ reads:
\begin{eqnarray}
p_k&=&p_{k i}-\frac{\textbf{p}_{\bot i}\cdot\textbf{A}(\phi_i)-A(\phi_i)^2/2}{c},
\label{pkpi}\\
p_k&=&p_{k i}-\frac{\textbf{p}_{\bot }\cdot\textbf{A}(\phi_i)+A(\phi_i)^2/2}{c}.
 \end{eqnarray}
In the quasistatic regime and neglecting the sub-barrier nondipole dynamics, $p_{k i}=0$ and $\textbf{p}_{\bot i}=0$, and the peak of the final momentum distribution within the  Simpleman model is:
\begin{eqnarray}
\textbf{p}_{\bot}^{(m)}(\phi_i)&=&-\textbf{A}(\phi_i) \label{pxm}\\
p_{k}^{(m)}(\phi_i)&=& \frac{p_{\bot}^{(m)}(\phi_i)^2}{2c}.
\label{pzm}
\end{eqnarray}
We define the light-front momentum via the integral of motion in a plane wave $p_- =c(1-\Lambda)$:
\begin{equation}
\label{p---}
p_-= p_{k }- \frac{\mathbf{p}_{\bot }^2}{2c}.
\end{equation}
In the quasistatic Simpleman picture the most probable value of the light-front momentum is, therefore, vanishing
\begin{equation}
\label{p-m}
  p_-^{(m)}(\phi_i)=0.
\end{equation}
The relationship of Eq.~(\ref{p-m}) for the time-resolved light-front momentum is fulfilled in a plain wave laser field of any  intensity and ellipticity as far as nonadiabatic and Coulomb effects, sub-barrier nondipole effects, as well as  recollisions are negligible. For this reason the momentum variable of $p_-(\phi_i)$ is a very convenient observable for the time-resolved investigation of  signatures of nonadiabatic, sub-barrier, and Coulomb effects. Note that recollisions do not play a significant role at  rather large ellipticity of the laser field $\epsilon\gtrsim 0.3$ \cite{Liu_2012,Maurer_2018}.

\subsection{Sub-barrier corrections}

In this section we  improve the Simpleman model  including the sub-barrier nondipole, Coulomb, and nonadiabatic corrections.
The sub-barrier nondipole effects shift the peak of the longitudinal momentum distribution  at the tunnel exit from the Simpleman value $p_{ki}=0$ to:
\begin{eqnarray}
p_{ki}=\frac{I_p}{3c}\left[1+ 6\nu\frac{E(\phi_i)}{E_a} \right],
 \end{eqnarray}
where the first term $I_p/(3c)$ is due to the sub-barrier nondipole magnetic field effect \cite{Klaiber_2013}, and the second term due to  the sub-barrier Coulomb field effect in the quasistatic and quasiclassical approximation \cite{He_2021}, $\nu$ is the effective principal quantum number of the bound state, and $E_a=\kappa^3$  the atomic field strength.

In the nonadiabatic regime the peak of the transverse distribution in the polarization plane at the tunnel exit is shifted due to the action of the nonadiabatic transverse force $F_\bot \sim E'(\phi_i)\tau_K\sim \epsilon \gamma(\phi_i) E(\phi_i)$ with respect to the direction of the tunneling channel during the sub-barrier dynamics within the Keldysh time $\tau_K=\gamma(\phi_i)/\omega$, with the Keldysh parameter $\gamma(\phi_i)=\omega \kappa/E(\phi_i)$.  This yields a transverse nonadiabatic momentum shift~\cite{Klaiber_2015}
\begin{eqnarray}
\mathbf{p}_{\bot i}^{(nad)}=\frac{\epsilon\gamma (\phi_i)\kappa}{6}\hat{\mathbf{e}}_\bot(\phi_i),
  \label{nonad-exit}
\end{eqnarray}
where
\begin{equation}\label{e-unit}
  \hat{\mathbf{e}}_\bot(\phi_i)=( E_y(\phi_i) ,- E_x(\phi_i))/E(\phi_i),
\end{equation}
is the unit vector perpendicular to the time-dependent laser field.

\subsection{Coulomb corrections in the continuum}

During the continuum motion of the ionized electron, the Coulomb field of the atomic core induces a momentum transfer. In the case of large ellipticity $\epsilon \gtrsim 0.3$, recollisions are negligible and the Coulomb effect  mostly arises during the electron motion near the tunnel exit with the coordinate $\mathbf{r}_e(\phi_i)=-I_p\mathbf{E}(\phi_i)/E(\phi_i)^2$.

While in the quasistatic limit the CCs at the tunnel exit are known \cite{Shvetsov-Shilovski_2009,Danek-II_2018}, here we derive the CC including nonadiabatic effects. The CC to the momentum due to the atomic potential $V(\mathbf{r} )$ is calculated as follows
\begin{eqnarray}
 \delta \mathbf{p}_C=-\int_{\eta_i}^\infty d\eta \nabla V(\mathbf{r}(\eta,\eta_i)),
\end{eqnarray}
using the electron trajectory $\mathbf{r}(\eta,\eta_i)$ in the laser field. The following results are obtained.

The CC to the momentum in the field direction reads:
\begin{eqnarray}
\label{pcE}
\delta \mathbf{p}_{\mathbf{e}C}=\pi  Z \frac{\mathbf{E}(\phi_i)}{E_a}
\left[1+\frac{\gamma(\phi_i)}{3\pi}\frac{1-\epsilon^2}{1+\epsilon^2}\frac{E_0^2\sin(2\phi_i)}{E(\phi_i)^2} \right],
\end{eqnarray}
where the first term coincides with the  quasistatic Coulomb momentum transfer  in the field direction  derived in \cite{Shvetsov-Shilovski_2009}, which results in a rotation of the final PMD  in the polarization plane, and inducing the attoclock offset angle  $\delta \theta\sim \pi \omega Z/(\epsilon \kappa^3)$ with respect to the Simpleman most probable angle $\theta_0=\pi/2$. The second term $\sim \gamma (\phi_i)$ is the nonadiabatic CC.

The CC to the momentum in the polarization plane, transverse to the field direction is:
\begin{eqnarray}
 \label{CC-tr-nad}
\delta \mathbf{p}_{\mathbf{e}_\bot C}=-\frac{\epsilon \gamma (\phi_i) \kappa}{6} \frac{2Z}{\kappa}\frac{E(\phi_i)}{E_a} \left[1+\frac{E_0^2}{E(\phi_i)^2(1+\epsilon^2)} \right]\hat{\mathbf{e}}_\bot(\phi_i)\nonumber\\
\end{eqnarray}
The first term is the CC due to the initial transverse nonadiabatic momentum following from \cite{Shvetsov-Shilovski_2009}. The second term  is an additional  CC in the continuum due to the motion driven by the nonadiabatic transverse force $\delta p_\bot \sim \epsilon\gamma(\phi_i) \kappa$, see intuitive explanation in Appendix \ref{A1}.

The CC to the momentum in the laser propagation direction:
\begin{eqnarray}
\label{pck}
\delta p_{kC}&=&- \left(p_{ki}+\frac{I_p }{3c}\right) \frac{2Z}{\kappa} \frac{E(\phi_i)}{E_a} \nonumber\\ &-&\frac{3\pi}{16}\gamma(\phi_i)\frac{I_p}{3c}\frac{Z}{\kappa}\frac{E(\phi_i)}{E_a} \frac{1-\epsilon^2}{1+\epsilon^2}\frac{E_0^2 \sin(2\phi_i)}{E(\phi_i)^2} ,
\end{eqnarray}
where the first term with the factor $p_{ki}$ is the  quasistatic Coulomb momentum transfer  in the direction transverse to the field  derived in \cite{Shvetsov-Shilovski_2009}, while the second term with the factor $I_p/(3c)$ is due to the electron nondipole displacement in the continuum by the $v\times B$  force, see  the intuitive explanation in Appendix \ref{A2}, and the last term is the nonadiabatic CC.

Thus, taking into account the  nondipole, Coulomb, and nonadiabatic effects under-the-barrier and in the continuum, we have
 \begin{eqnarray}
 \mathbf{p}_{\bot i}&=& \pi  Z \frac{\mathbf{E}(\phi_i)}{E_a}\left[1+g_{\mathbf{e}}(\phi_i) \right]+\frac{\epsilon\gamma (\phi_i)\kappa}{6}\hat{\mathbf{e}}_\bot(\phi_i)\left[ 1-g_\bot(\phi_i)\right],\nonumber\\ \label{pbotCgi} \\
 p_{ki}&=&\frac{I_p}{3c}\left[ 1+6\nu\frac{E(\phi_i)}{E_a}-g_k(\phi_i)\right] \label{pkCgi},
\end{eqnarray}
with the nonadiabatic and Coulomb correction functions
\begin{eqnarray}
g_{\mathbf{e}}(\phi_i) &=& \frac{\gamma(\phi_i)}{3\pi}\frac{1-\epsilon^2}{1+\epsilon^2}\frac{E_0^2\sin(2\phi_i)}{E(\phi_i)^2}, \label{g_E}\\
g_\bot (\phi_i)&=& \frac{2Z}{\kappa}\frac{E(\phi_i)}{E_a}\left(1+\frac{E_0^2}{E(\phi_i)^2(1+\epsilon^2)} \right),\label{g_bot}\\
g_k (\phi_i) &=&\frac{E(\phi_i)}{E_a}\left[ \frac{4Z}{\kappa}+\frac{3\pi}{16}\gamma(\phi_i)\frac{Z}{\kappa}\frac{1-\epsilon^2}{1+\epsilon^2}\frac{E_0^2 \sin(2\phi_i)}{E(\phi_i)^2}\right]. \label{g_k}
\end{eqnarray}
Here we keep the leading terms in $E_0/E_a$, and have added the negative continuum Coulomb corrections to the initial momentum, assuming that it takes place during the motion near the tunnel exit in the case of a large ellipticity.
With Eqs.~(\ref{pbotCgi}),(\ref{pkCgi})  the most probable asymptotic momentum reads
\begin{eqnarray}
 \mathbf{p}_{\bot }^{(m)}(\phi_i)&=& -\mathbf{A}(\phi_i)+\pi  Z\frac{\mathbf{E}(\phi_i)}{E_a}\left[1+g_{\mathbf{e}}(\phi_i) \right]\nonumber\\
 &+&\frac{\epsilon\gamma (\phi_i)\kappa}{6}\hat{\mathbf{e}}_\bot(\phi_i)\left[ 1-g_\bot(\phi_i)\right], \label{pbotCg}\\
  p_{k}^{(m)}(\phi_i)&=&\frac{I_p}{3c}\left[ 1+6\nu\frac{E(\phi_i)}{E_a}-g_k(\phi_i)\right]+\frac{\mathbf{A}(\phi_i)^2}{2c}\nonumber\\
  &-&\frac{\pi  Z}{E_a} \frac{\mathbf{E}(\phi_i)\cdot\mathbf{A}(\phi_i)}{c}\left[1+g_{\mathbf{e}}(\phi_i) \right]\label{pkCg}\\
  &-&\frac{\epsilon\gamma (\phi_i)\kappa}{6}\frac{\hat{\mathbf{e}}_\bot(\phi_i)\cdot\mathbf{A}(\phi_i)}{c}\left[ 1-g_\bot(\phi_i)\right]\nonumber,
\end{eqnarray}
The attoclock angle is defined as
  \begin{eqnarray}
  \label{attoclock-angle}
\tan \theta(\phi_i)=p_y( \phi_i )/p_x( \phi_i ),
\end{eqnarray}
which provides a mapping of the initial laser phase of the tunneled electron to the attoclock angle.

The light front momentum  Eq.~(\ref{p---}) is an integral of motion:
\begin{eqnarray}
\label{p-minus}
p_{-}(\eta)=p_{k }(\eta)-\frac{p_{\bot }^2(\eta)}{2c}=p_{k i }-\frac{p_{\bot i}^2}{2c}.
 \end{eqnarray}
From the latter, keeping the first order terms with respect to $E_0/E_a$ and $\gamma$, we have for the peak value of the asymptotic light-front momentum:
\begin{eqnarray}
 p_{-}^{(m)}(\phi_i)= \frac{I_p}{3c}\left[ 1+6\nu\frac{E(\phi_i)}{E_a}-g_k(\phi_i)\right].
\label{p-minus-imroved}
\end{eqnarray}
The term $ p_{\bot i}^2/(2c)$ in Eq.~(\ref{p-minus}) has contributions of the order of magnitude $O((E_0/E_a)^2, \gamma^2)$, which are neglected.

Thus, we have derived in the weakly nonadiabatic regime the most probable asymptotic momentum of the photoelectron within the Simpleman model [Eqs.~(\ref{pbotCg}),(\ref{p-minus-imroved})], which provides the parametric dependence of the asymptotic momentum on the attoclock angle $\theta$ via the parameter $\phi_i$ [Eq.~(\ref{attoclock-angle})]. The estimation for the light-front momentum Eq.~(\ref{p-minus-imroved}) includes the nondipole sub-barrier momentum shift [$I_p/3c$],  quasistatic CC during the sub-barrier dynamics [$6\nu E(\phi_i)/E_a$] and in the continuum [$(-4Z/\kappa)  E(\phi_i)/E_a$], as well as nonadiabatic CC [$\sim \gamma(\phi_i)2Z/\kappa$]. The estimation for the transverse momentum Eq.~(\ref{pbotCg}) includes the  quasistatic CC during the  continuum dynamics [$\pi Z  E(\phi_i)/E_a$], and its nonadiabatic correction [$\sim \gamma(\phi_i)$] as well as the nonadiabatic momentum shift due to sub-barrier dynamics [$\epsilon \gamma(\phi_i)\kappa/6$], and its CC [$\sim (2Z/\kappa) (E(\phi_i)/E_a)$].

\begin{figure*}
   \begin{center}
   \includegraphics[width=0.4\textwidth]{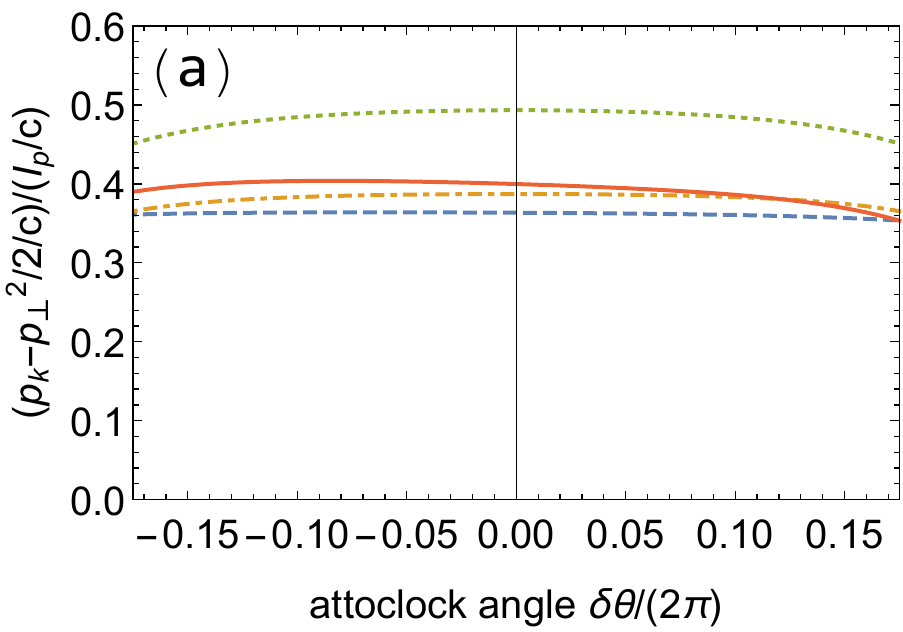}
   \includegraphics[width=0.4\textwidth]{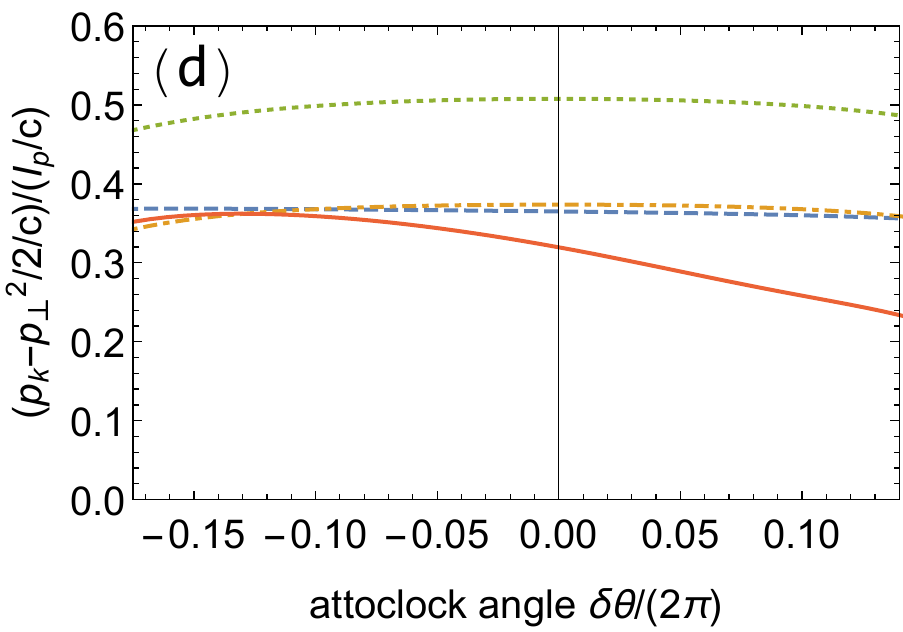}
 \includegraphics[width=0.4\textwidth]{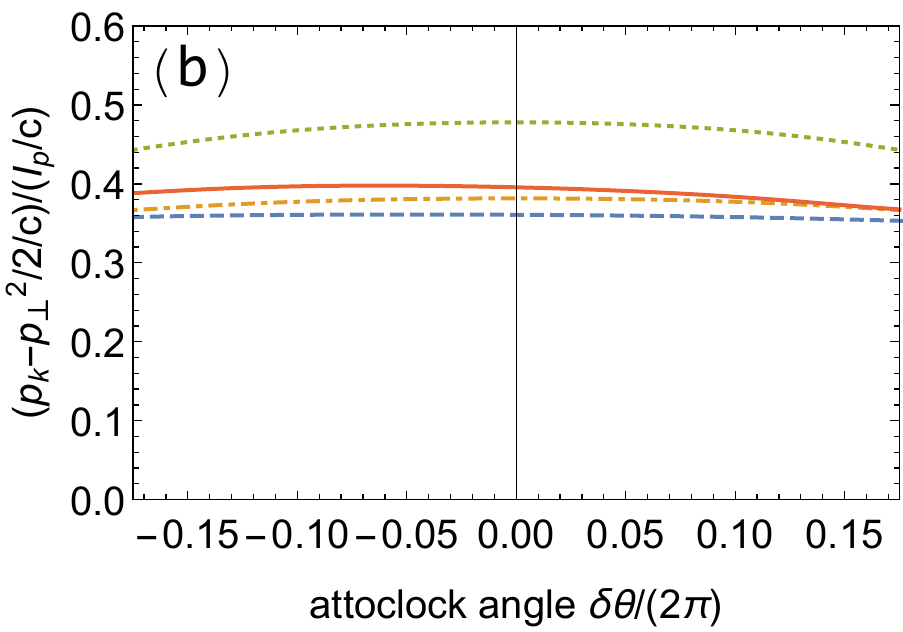}
   \includegraphics[width=0.4\textwidth]{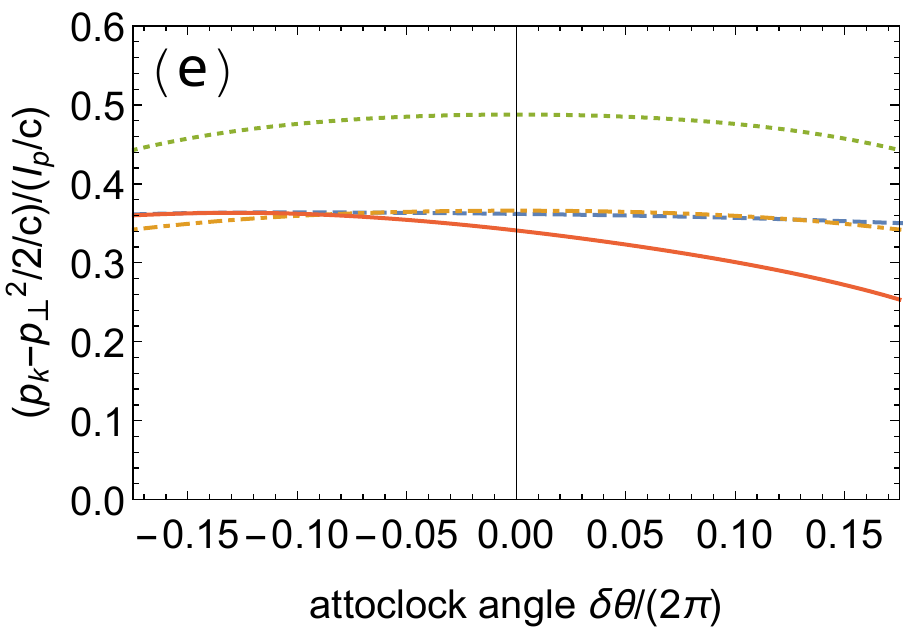}
\includegraphics[width=0.4\textwidth]{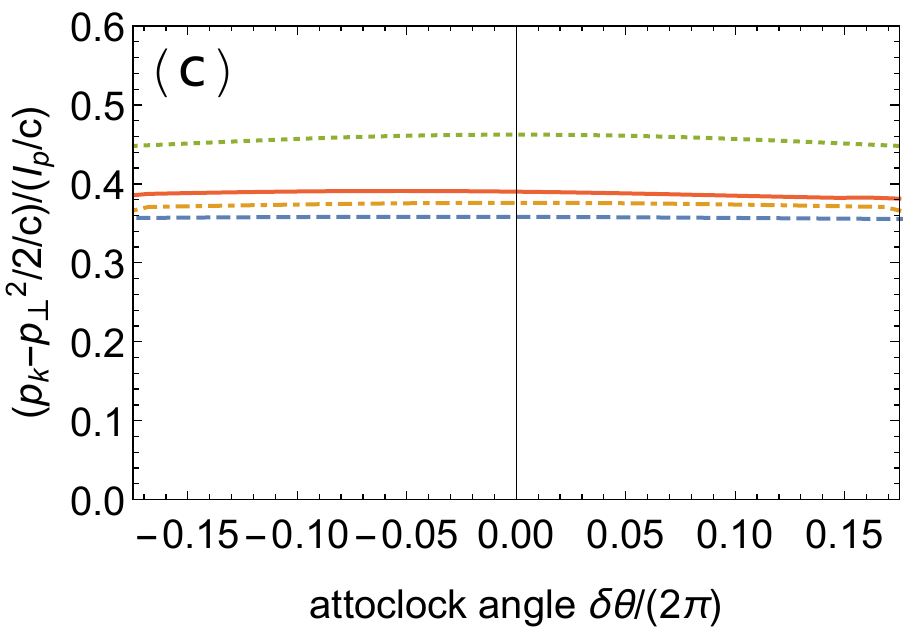}
 \includegraphics[width=0.4\textwidth]{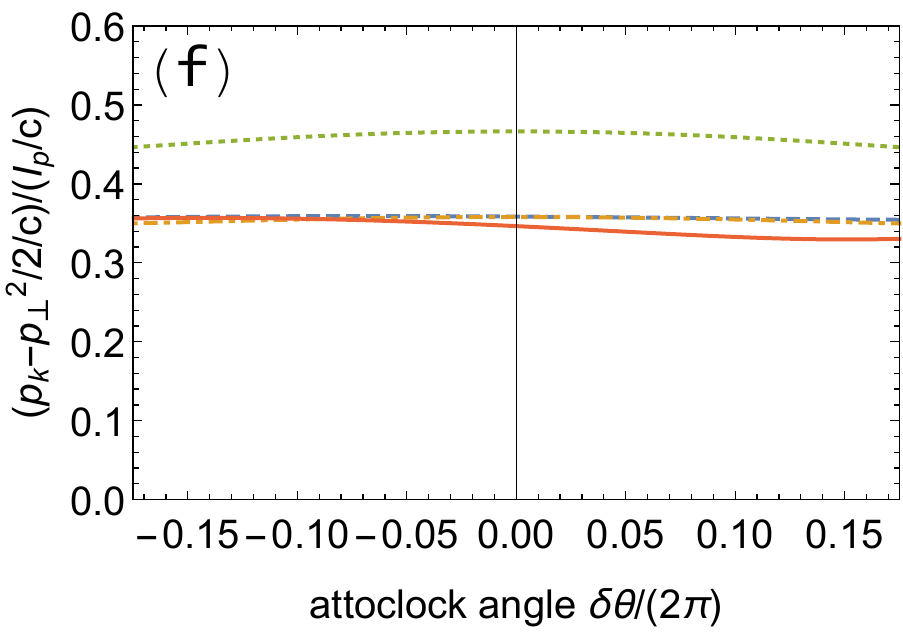}
\caption{Light-front momentum $p_-=p_{k }- \frac{p_{\bot }^2}{2c}$ vs  attoclock offset angle $\delta \theta$: Left column - quasistatic regime, $\omega=0.02$ ($\gamma\approx 0.4$); Right column - nonadiabatic regime $\omega=0.05$ ($\gamma\approx 1.1$); for ellipticity values (a,d) $\epsilon=0.5$, (b,e) $\epsilon=0.7$, (c,f) $0.9$; (red-solid)  CCSFA, (orange-dash-dotted) plain SFA without CC, (green-dotted) TCSFA (SFA with only sub-barrier CC), (blue-dashed) improved Simpleman model. The laser field strength  is $E_0=0.05$.  }
       \label{fig1}
    \end{center}
  \end{figure*}

\begin{figure}
\begin{center}
\includegraphics[width=0.4\textwidth]{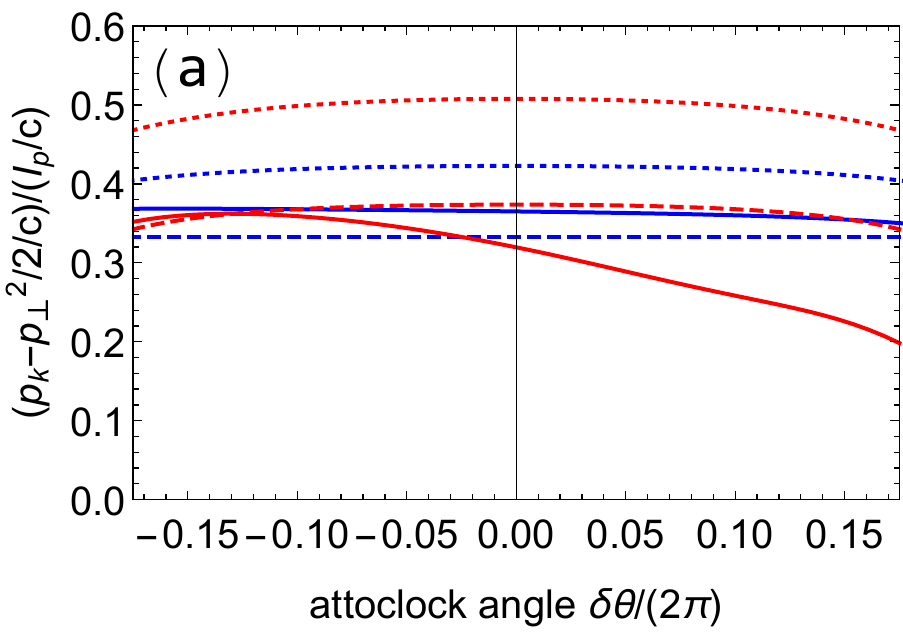}
\includegraphics[width=0.37\textwidth]{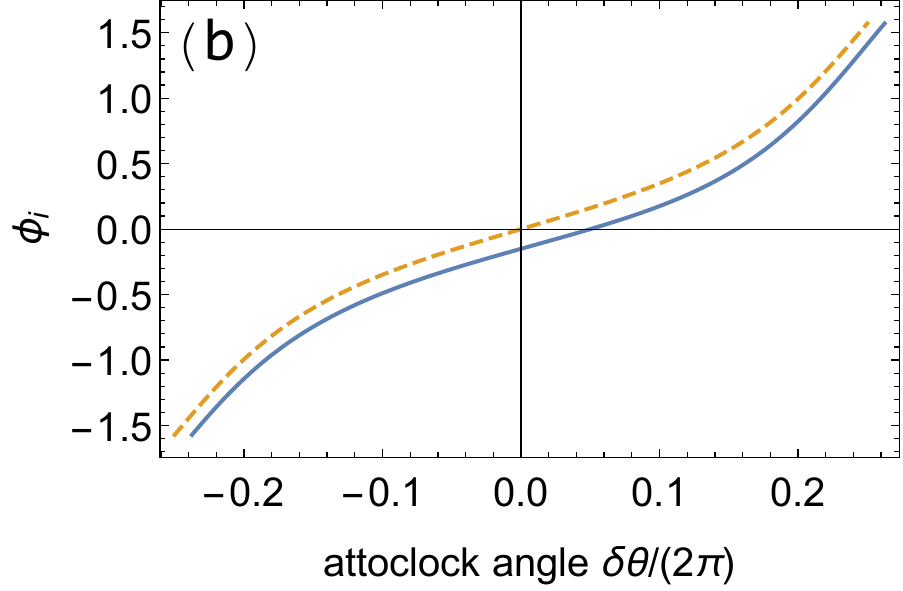}
\caption{(a) Different contributions to the Simpleman model for the nonadiabatic regime, with $\omega=0.05$ and ellipticity $\epsilon=0.5$: (red-dashed) plain SFA with no CC, (blue-dashed) Simpleman with no CC, (red-dotted) TCSFA (no continuum CC), (blue-dotted) Simpleman with no continuum  CC, (red, solid)  CCSFA, (blue-solid) improved Simpleman model with all corrections; (b) Ionization phase vs attoclock offset angle $\delta \theta$ via the improved Simpleman model Eq.~(\ref{attoclock-angle}) (blue, solid), plain Simpleman (orange, dashed). The laser field strength  is $E_0=0.05$. }
\label{fig2}
\end{center}
\end{figure}
\begin{figure*}
\begin{center}
\includegraphics[width=0.4\textwidth]{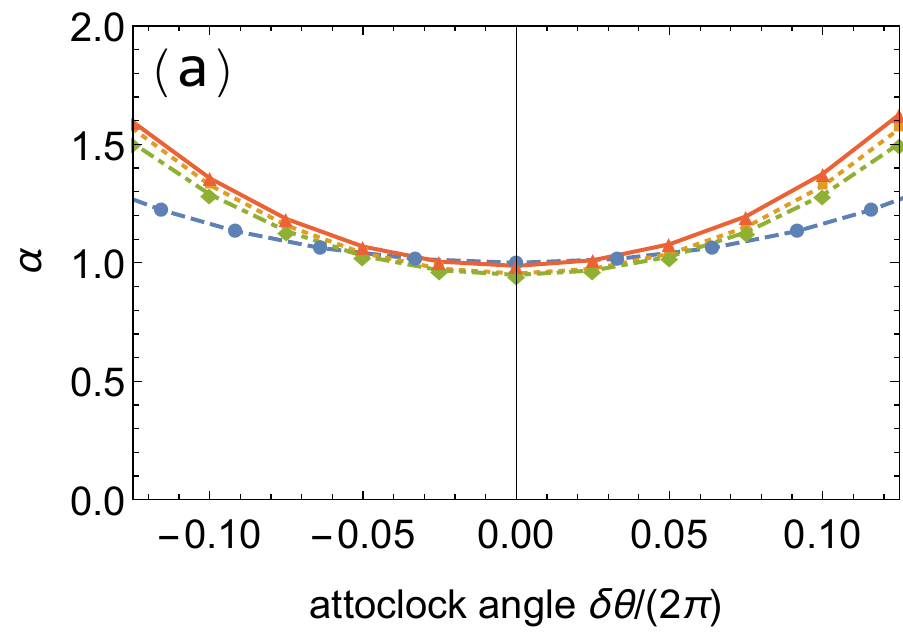}
\includegraphics[width=0.4\textwidth]{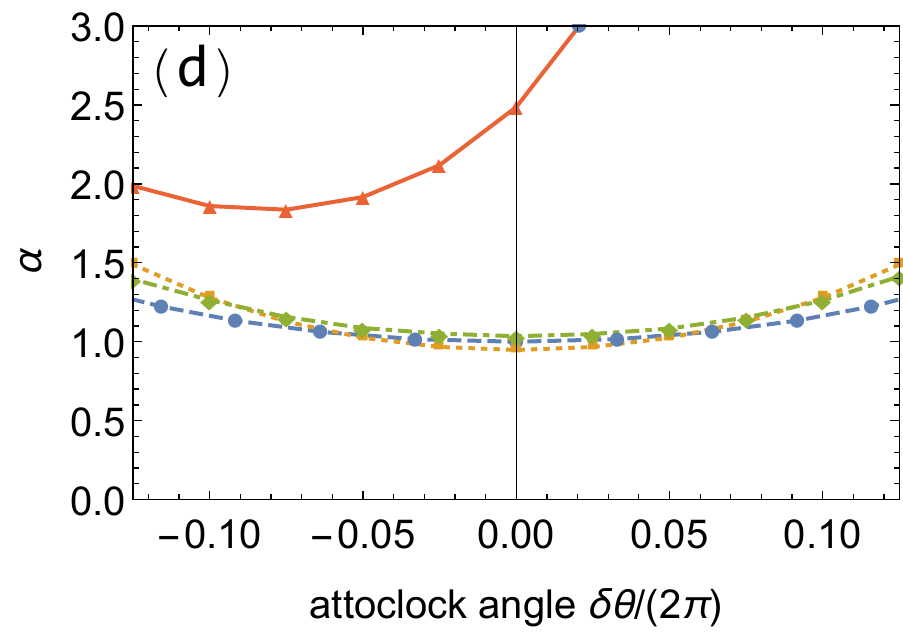}
\includegraphics[width=0.4\textwidth]{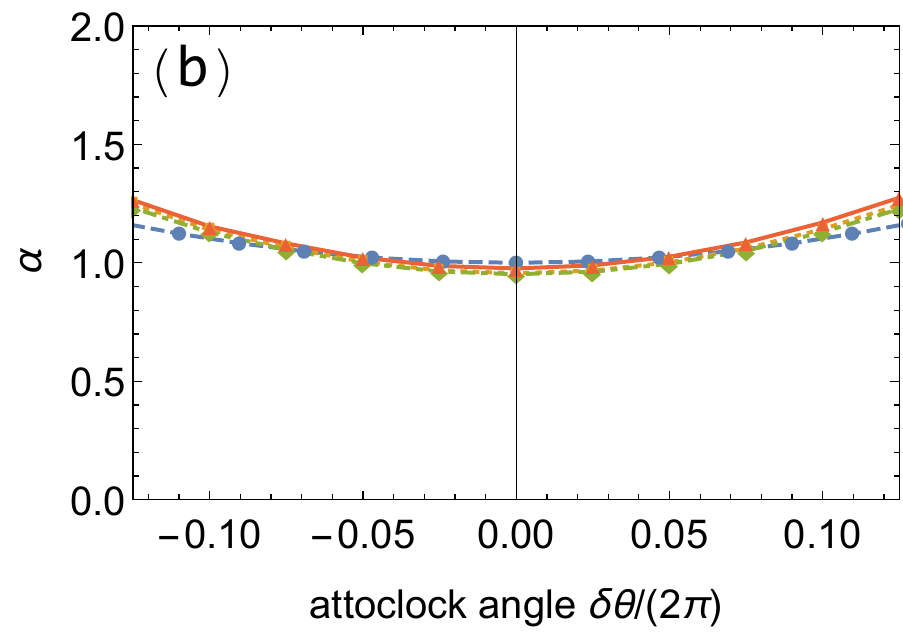}
\includegraphics[width=0.4\textwidth]{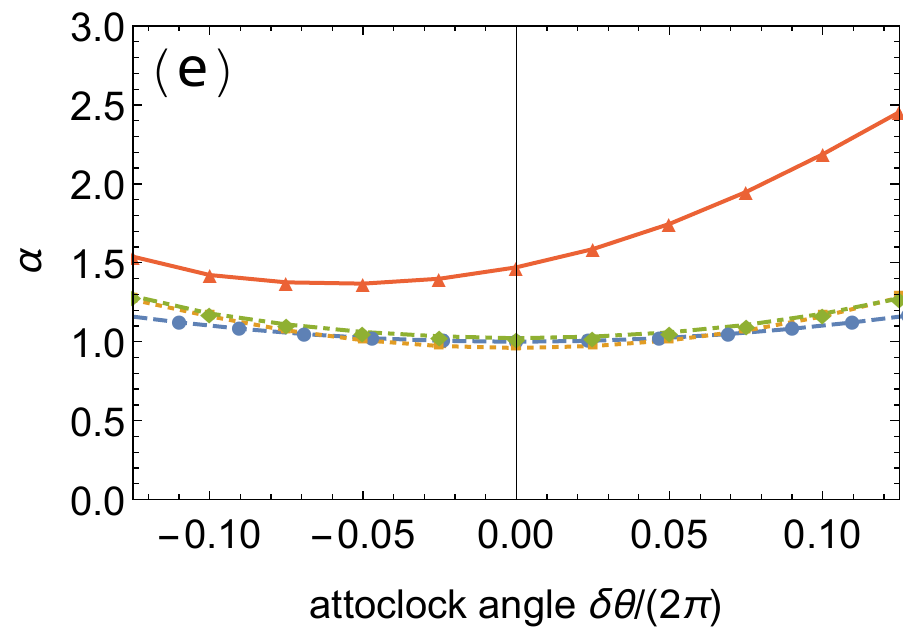}
\includegraphics[width=0.4\textwidth]{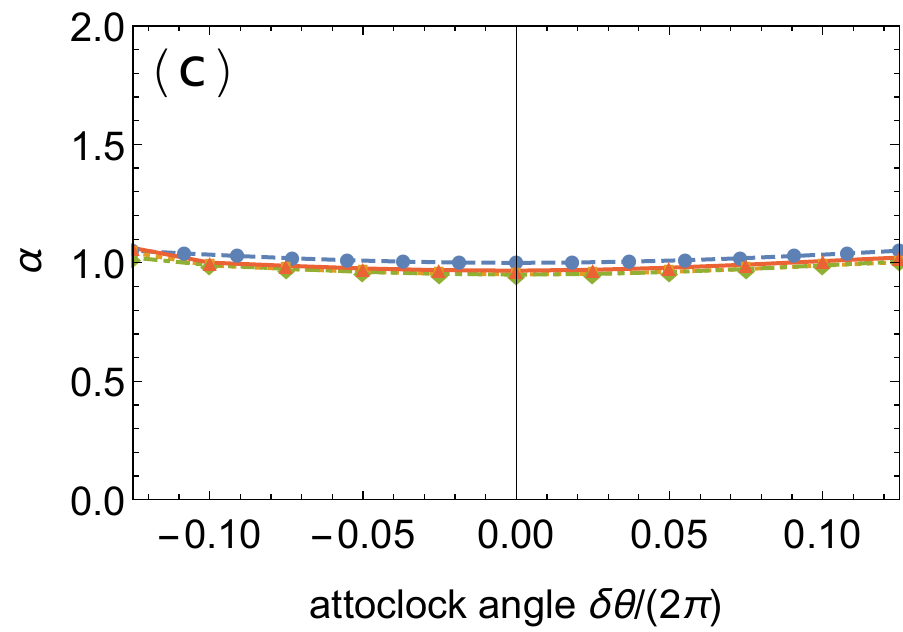}
\includegraphics[width=0.4\textwidth]{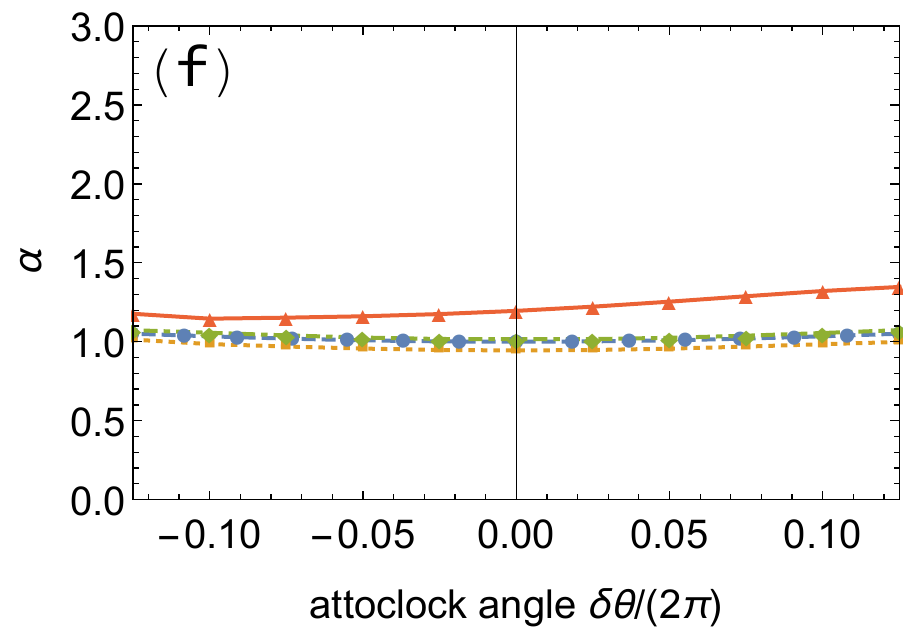}
\caption{  The coefficient $\alpha$ from Eq.~(\ref{pbotmaxtheta}) vs attoclock offset angle $\delta \theta$.
Left column - quasistatic regime, $\omega=0.02$ ($\gamma\approx 0.4$); Right column - nonadiabatic regime $\omega=0.05$ ($\gamma\approx 1.1$); for ellipticity values (a,d) $0.5$, (b,e) $0.7$, (c,f) $\epsilon=0.9$; The laser field strength  is $E_0=0.05$. }
\label{fig3}
\end{center}
\end{figure*}
\begin{figure}
\begin{center}
\includegraphics[width=0.4\textwidth]{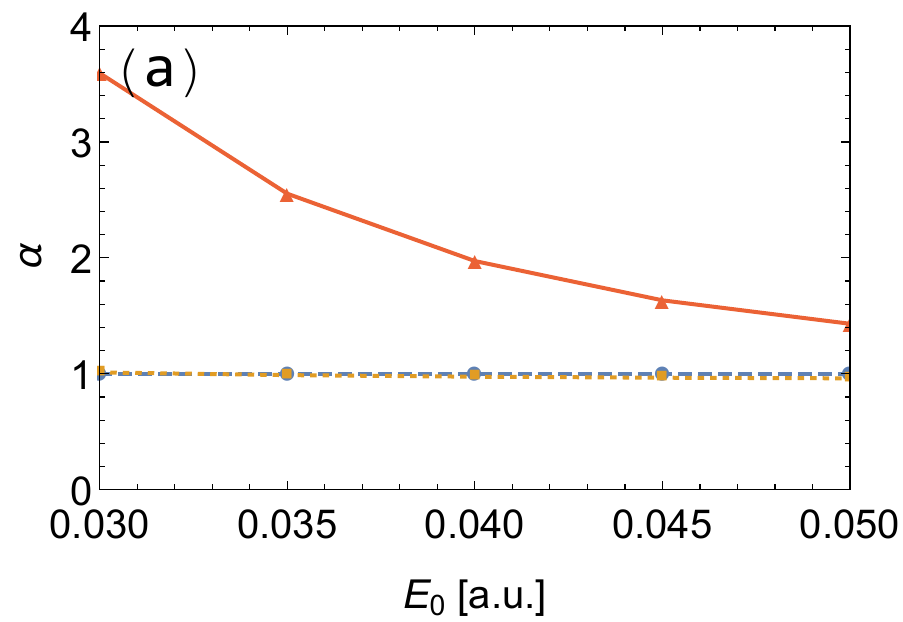}
\includegraphics[width=0.4\textwidth]{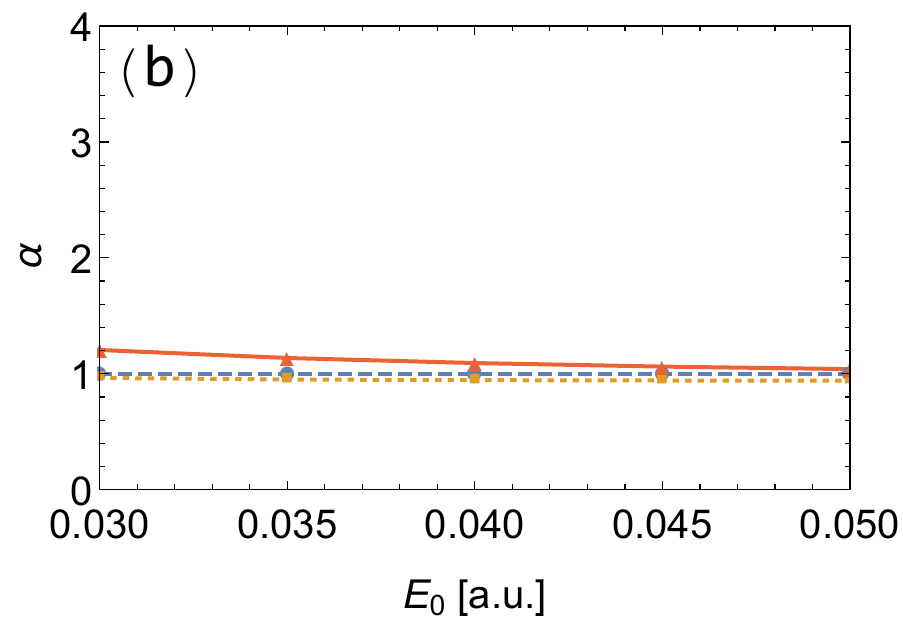}
\caption{  The coefficient $\alpha$ [Eq.~(\ref{pbotmaxtheta})]  vs the laser field, for $\omega=0.04$ a.u. and $\kappa=1$ a.u.: (a) $\epsilon =0.6$; (b) $\epsilon =1$;  (red-solid)  CCSFA, (orange-dash-dotted) plain SFA without CC, (blue-dashed) improved Simpleman model.  }
\label{fig4}
\end{center}
\end{figure}

\begin{figure}[b]
   \begin{center}
 \includegraphics[width=0.4\textwidth]{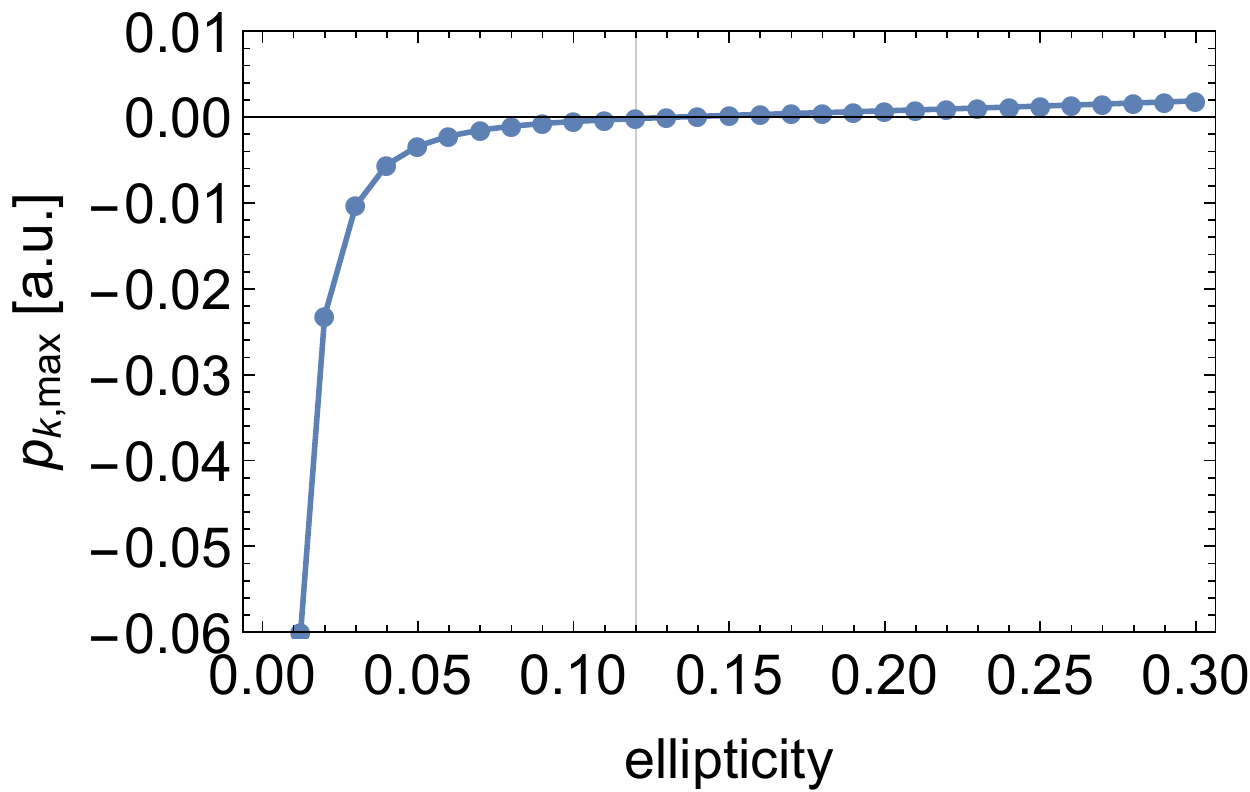}
              \caption{The maximum of the longitudinal momentum via the improved Simpleman (with forward rescattering) vs ellipticity,  for $\omega=0.01345$ a.u., $\kappa=0.944$ a.u. and $E_0=0.0338$ a.u. }
       \label{fig5}
    \end{center}
  \end{figure}
  \begin{figure}[b]
   \begin{center}
 \includegraphics[width=0.4\textwidth]{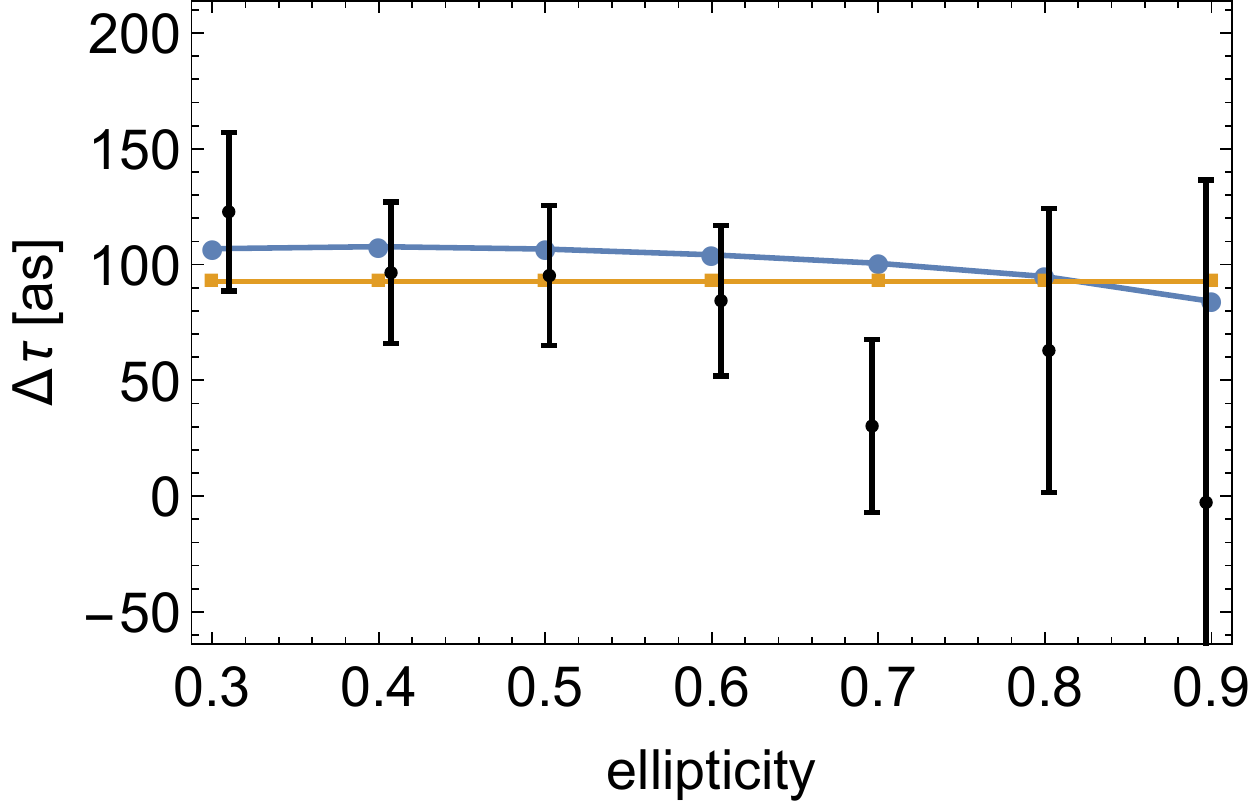}
             \caption{Offset angle (expressed as a time delay) between the attoclock angle of the maximum yield and the minimum of the longitudinal momentum vs ellipticity.   Experimental data \cite{Willenberg_2019} with error bars are black, CCSFA - blue cycles, and Simpleman results - orange boxes. }
       \label{fig6}
    \end{center}
  \end{figure}
\begin{figure*}
   \begin{center}
 \includegraphics[width=0.4\textwidth]{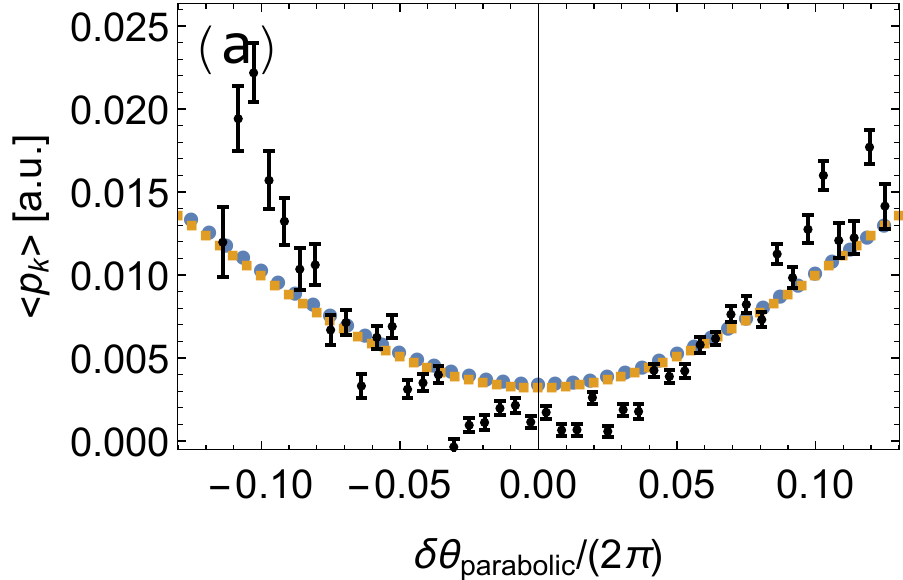}
 \includegraphics[width=0.4\textwidth]{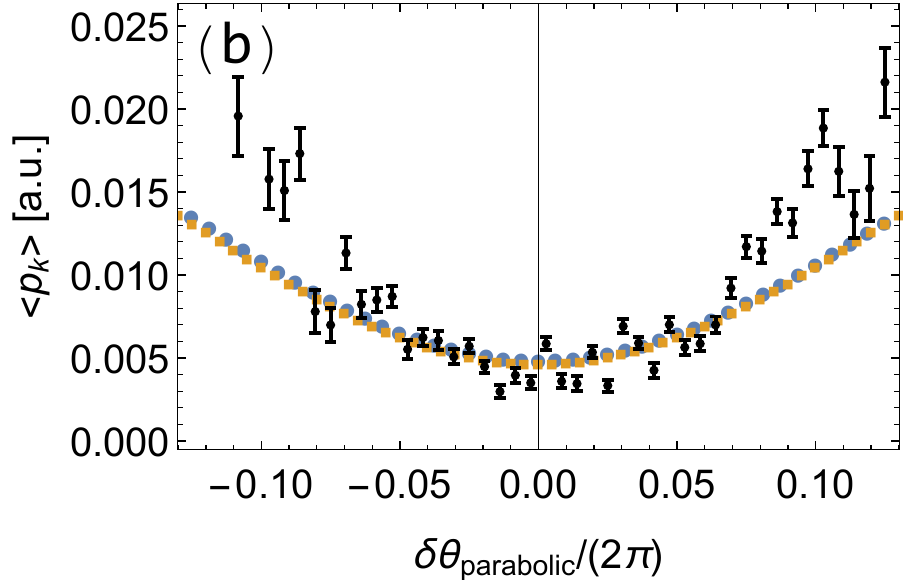}
 \includegraphics[width=0.4\textwidth]{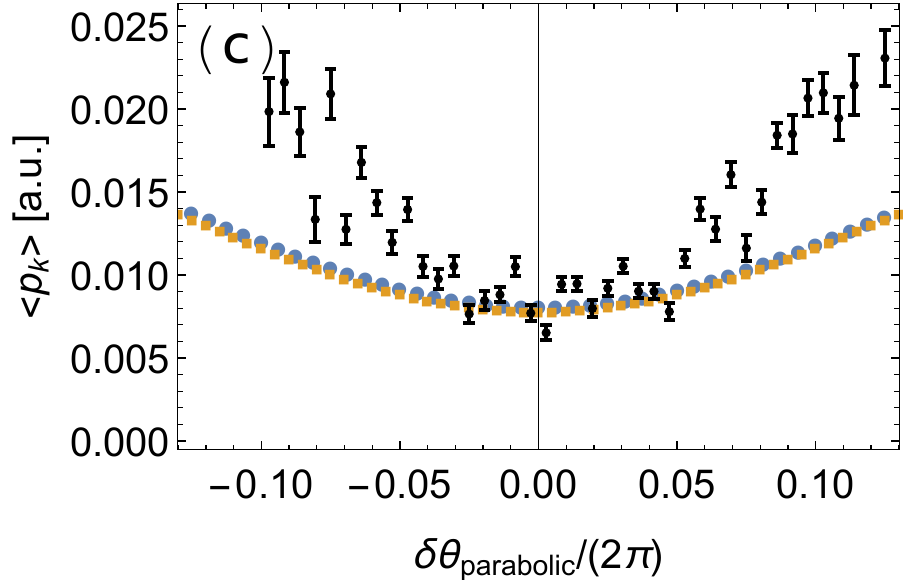}
 \includegraphics[width=0.4\textwidth]{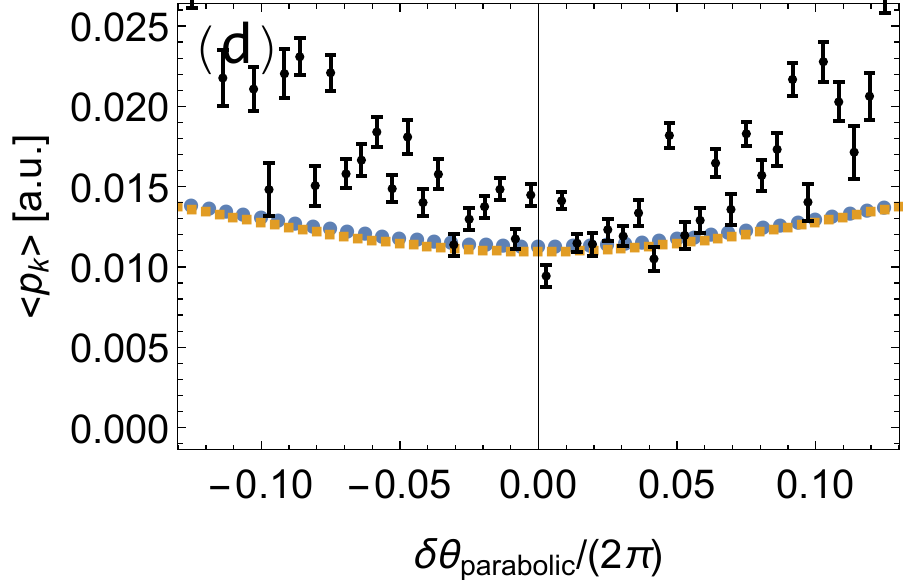}
           \caption{Longitudinal momentum vs attoclock offset angle: (a) $\epsilon=0.3$; (b) $\epsilon=0.4$; (c) $\epsilon=0.6$; (d) $\epsilon=0.8$.  Experimental data \cite{Willenberg_2019} with error bars are black, CCSFA - blue cycles, and Simpleman results - orange boxes. The calculations show the average $p_k$ for the given offset angle. }
       \label{fig7}
    \end{center}
  \end{figure*}
\begin{figure} [b]
   \begin{center}
 \includegraphics[width=0.45\textwidth]{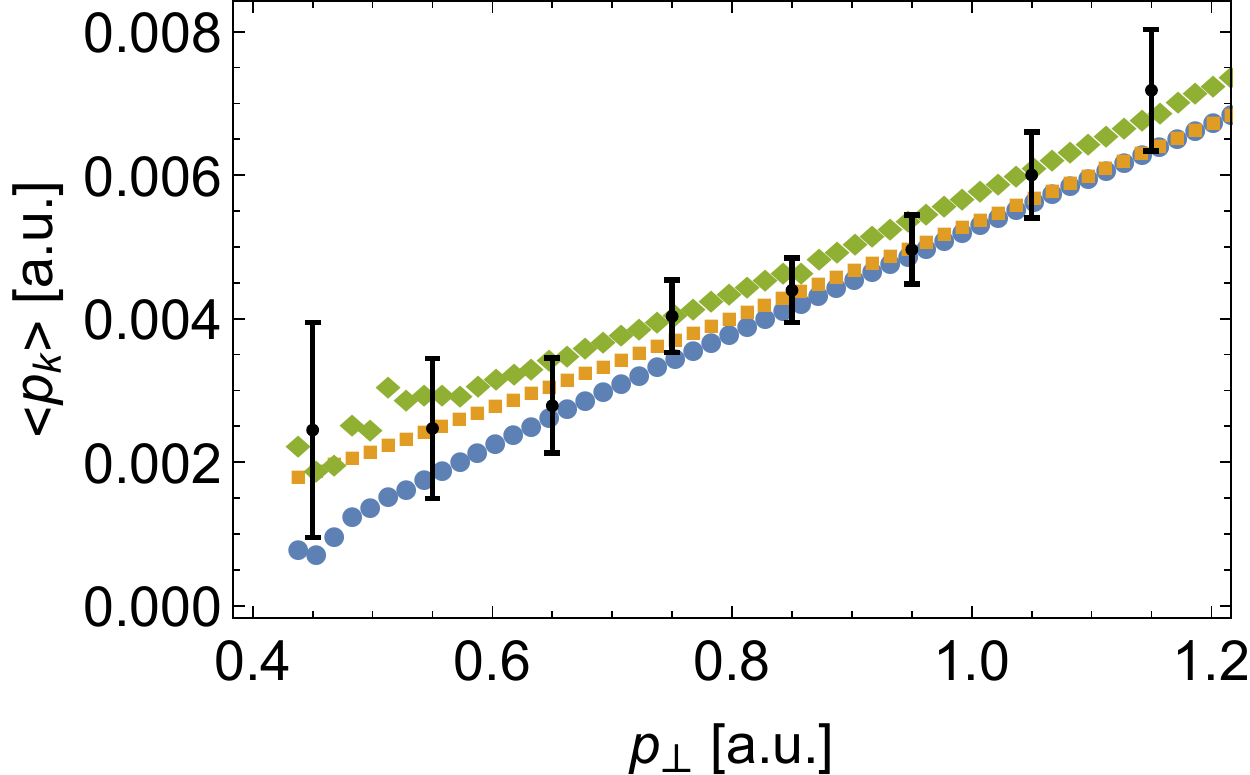}
              \caption{The average of the longitudinal momentum vs the transverse momentum:  CCSFA (blue dots), plain SFA (orange dots),  CCSFA with only sub-barrier corrections (green dots), and the experimental data (black points) of Ref.~\cite{Hartung_2019}.}
       \label{fig8}
    \end{center}
  \end{figure}

\section{Comparison of SFA results with improved Simpleman model}
\subsection{Time-resolved light-front momentum}

In this section we consider the ellipticity dependence of the time-resolved light-front momentum $p_- (\phi)$ in  the quasistatic and nonadiabatic regimes, respectively, see Fig.~\ref{fig1}. The use of the light-front momentum for the presentation of the results is quite useful, because it immediately demonstrates the role of Coulomb and nonadiabatic corrections, as in the plain Simpleman model $p_-=0$. We have applied several versions of SFA: 1) full CCSFA, which include all Coulomb corrections, i.e.  during the sub-barrier dynamics, as well as in the continuum; 2) Tunnel-Coulomb-corrected SFA (TCSFA), the SFA with only sub-barrier Coulomb corrections; 3) plain SFA with no Coulomb corrections. We provide also a comparison of the SFA results with the improved Simpleman model [Eq.~(\ref{p-minus-imroved})].

The general observation from the results of Fig.~\ref{fig1} is the following. In the quasistatic regime (in our example $\gamma\approx 0.4$), the improved Simpleman model and the plain SFA describe quite well the full CCSFA results for the given ellipticity range $\epsilon=0.5-0.9$. They both  underestimate the CCSFA result for $p_-$ slightly. The deviation of CCSFA result from the improved Simpleman model and the plain SFA  is not large, because the sub-barrier CC (highlighted via TCSFA) and the continuum CC (included in CCSFA), which are of opposite sign and larger at small ellipticity values,  compensate each other to some extent \cite{He_2021}.

In the nonadiabatic regime [in our example $\gamma\approx 1.1$, Fig.~\ref{fig1} (right column)] there are large deviations of the plain SFA with respect to CCSFA at small ellipticity. The performance of the improved Simpleman model is also not good. It doesn't predict the slope the CCSFA result, there is a deviation from CCSFA in the offset angle dependence. The deviation is larger at small ellipticities and at large positive offset angles. The improved Simpleman model does not capture this latter feature. This  stems from  nonadiabatic Coulomb corrections, which are larger for small elipticity. In the nonadiabatic regime the electron stays longer near the core than the quasistatic estimation assumes, leading to a large CC. In the Simpleman model we expand the nonadiabatic CC with respect to $\gamma$, and the model is not accurate at large $\gamma$.  The characteristic feature of the CC in  the nonadiabatic regime is that it induces an asymmetry between the positive and negative offset angles. At $\gamma \sim 1$  this effect is significant.

Contribution of different Coulomb and nonadiabatic corrections are analyzed in Fig.~\ref{fig2}(a) for the nonadiabatic regime. The  Simpleman model without CC (but with nonadiabatic corrections) coincides with the plain SFA result for the time-resolved light-front momentum.  The sub-barrier and continuum CCs are underestimated by the Simpleman model, because of the applied expansion over $\gamma$-parameter (weakly nonadiabatic approximation).  This results in the final deviation of the improved Simpleman with respect to CCSFA. Especially the asymmetry of  $p_-$ with respect to the sign of the offset angle, which is due to the nonadiabatic Coulomb effects in the continuum are not captured in the Simpleman model. This asymmetry is also exhibited in Fig.~\ref{fig2}(b) showing the ionization phase with respect to the attoclock offset angle. Thus, the peak of the laser field $\phi_i=0$ is shifted from the zero offset angle, i.e. the field is not symmetric  with respect to $\delta \theta=0$.

\subsection{Transverse momentum distribution resolved in time and in longitudinal momentum}

In previous section we investigated the absolute peak value of the time-resolved light-front momentum [Eq.~(\ref{p-minus-imroved})]. Following the experiment \cite{Hartung_2021}, we further provide a more detailed description and examine the peak of the transverse momentum distribution resolved in the longitudinal momentum, as well as resolved in time (attoclock offset angle). For a given ionization phase $\phi_i$, let us fix $p_k$ and calculate the maximum of the transverse momentum distribution with respect to the transverse momentum $p_\bot$. The final distribution over transverse momenta arises because of the deviation of the electron transverse momentum at the tunnel exit from the peak value
\begin{eqnarray}
\mathbf{p}_{\bot e}&=&\hat{\mathbf{e}}_\bot(\phi_i)\left( \tilde{p}_\bot+\frac{\epsilon\gamma\kappa}{6} \right)\\
p_{ke}&=&\tilde{p}_k+I_p/(3c).
\end{eqnarray}
Note that $\mathbf{p}_{\bot\, i}=\mathbf{p}_{\bot e}-\delta \mathbf{p}_{\bot C}$ and $p_{ke}=p_{ki}-\delta p_{kC}$, with CCs $\delta \mathbf{p}_{\bot C}$ and $\delta p_{kC}$. Then from Eqs.~(\ref{pxpx}) and (\ref{pkpi}) we have
\begin{eqnarray}
     \mathbf{p}_{\bot}&=&\mathbf{p}_{\bot}^{(m)}(\phi_i)+\hat{\mathbf{e}}_\bot(\phi_i)\tilde{p}_{\bot}\left[ 1-g_\bot(\phi_i)\right]\\
     p_k&=&p_{k}^{(m)}(\phi_i)+\tilde{p}_k\left[ 1-g_k(\phi_i)\right]-\frac{\tilde{\mathbf{p}}_{\bot}\cdot\mathbf{A}(\phi_i)}{c}\left[ 1-g_\bot(\phi_i)\right],\nonumber\\ \label{pztilde}
  \end{eqnarray}
where $\mathbf{p}_{\bot }^{(m)}(\phi_i)$, $p_{k}^{(m)}(\phi_i)$ are the most probable asymptotic momentum components of the ionized wave packet at the tunnel exit via Eqs.~(\ref{pbotCg}),(\ref{pkCg}).
the factors $g_\bot (\phi_i)\,, g_k(\phi_i)$ account for the continuum CC due to the additional momentum $\tilde{p_k},\tilde{p}_\bot$.
The Eq.~(\ref{pztilde}) shows that $\tilde{p}_k$ and $\tilde{\mathbf{p}}_{\bot}$ are not independent at a given asymptotic momentum $p_k$.
\begin{eqnarray}
    \tilde{p}_\bot &=& \frac{\left[\mathbf{p}_{\bot}-\mathbf{p}_{\bot}^{(m)}(\phi_i)\right]\cdot \hat{\mathbf{p}}^{(m)}_\bot(\phi_i) }{\left[\hat{\mathbf{e}}_\bot(\phi_i)\cdot \hat{\mathbf{p}}^{(m)}_\bot(\phi_i)\right]\left[1-g_\bot(\phi_i) \right]}\label{pbottilde}\\
    \tilde{p}_k &=&\frac{p_k-p_k^{(m)}(\phi_i)}{1-g_k(\phi_i) }+\tilde{p}_\bot\frac{A_\bot(\phi_i)}{c}\frac{1-g_\bot(\phi_i)}{1-g_k(\phi_i)}\label{pktilde}
 \end{eqnarray}
where $A_\bot(\phi_i)=\mathbf{A}(\phi_i)\cdot \hat{\mathbf{e}}_\bot(\phi_i)$, and $\hat{\mathbf{p}}^{(m)}_\bot(\phi_i)\equiv \mathbf{p}_{\bot}^{(m)}(\phi_i)/|\mathbf{p}_{\bot}^{(m)}(\phi_i)|$.
The probability distribution over electron momenta at the tunnel exit $p_{ke}=\tilde{p}+I_p/(3c)$ and $p_{\bot e}=\tilde{p}_\bot+\epsilon\gamma\kappa/6$ is determined by the tunneling Perelomov-Popov-Terent'ev (PPT)-distribution in the nondipole and nonadiabatic regime \cite{PPT,ADK,Klaiber_2013_II}:
\begin{eqnarray}
w(\tilde{p}_{\bot},\tilde{p}_k)\propto \exp \left(-\frac{2}{3}\frac{\left(\kappa^2 +p_{\bot\,e}^2+ p_{k\,e}^2\right)^{3/2}}{E(\phi_i) }\left( 1-\frac{p_{k\,e}}{2c}\right) \left( 1-\frac{\epsilon\omega p_{\bot\,e}}{2E(\phi_i)}\right)\right)\nonumber\\
     \end{eqnarray}
Therefore, the maximum of the distribution for a given $p_k-p_k^{(m)}(\phi_i)$ is determined by the minimum of $G(p_{\bot e},p_{k\,e})$ [$w(p_{\bot e},p_{k e})\propto \exp\left(-G(p_{\bot e},p_{k e})\right)$]. From the condition $\partial G/\partial p_{\bot\,e}=0$, taking into account Eq.~(\ref{pktilde}),(\ref{pbottilde}), and keeping the terms linear in $p_{\bot e},p_{k e}$, and up to the order of $1/c$,  one obtains
 \begin{eqnarray}
 p_\bot-p_\bot^{(m)}=\alpha(\phi_i)
 \left[p_k(\phi_i)-p_k^{(m)}(\phi_i)\right]\frac{A_0}{c}, \label{pbotmax}
 \end{eqnarray}
with $p_\bot-p_\bot^{(m)}=[\mathbf{p}_{\bot } (\phi_i)-\mathbf{p}_{\bot}^{(m)}(\phi_i)]\cdot \hat{\mathbf{p}}_\bot^{(m)}(\phi_i)$, and the coefficient  $\alpha(\phi_i)$ defined as
 \begin{eqnarray}
 \alpha(\phi_i)\equiv -\frac{A_\bot(\phi_i) }{A_0 \left[\hat{\mathbf{e}}_\bot(\phi_i)\cdot \hat{\mathbf{p}}^{(m)}_\bot(\phi_i)\right]}\left[\frac{1-g_\bot(\phi_i)}{1-g_k(\phi_i)}\right] \label{alpha_phi_i},
\end{eqnarray}
and CC factors $g_\bot(\phi_i),g_k(\phi_i)$ from Eqs.(\ref{g_bot}), (\ref{g_k}).
Thus the time resolved PMD with respect to $(p_{\bot } (\phi_i),p_k(\phi_i))$ shows a local maximum, which runs along the line of Eq.~(\ref{pbotmax}). It is in accordance of the experimental qualitative observation of Ref.~\cite{Hartung_2021}: when $p_k>p_{k}^{(m)}$, one has a local maximum of $p_{\bot}$ at a given $p_k$, which exceeds the absolute maximum $p_{\bot}^{(m)}$, and vice verse. An alternative derivation of the correlation of the transverse and longitudinal components momentum is given in Appendix~\ref{A3}.

One may  define also the coefficient $\alpha$ as a function of the attoclock angle
 \begin{eqnarray}
    p_{\bot }(\theta)-p_{\bot}^{(m)}(\theta)=\alpha(\theta) \left[p_k(\theta)-p_{k}^{(m)}(\theta)\right]\frac{A_0}{c},
    \label{pbotmaxtheta}
\end{eqnarray}
where $p_{\bot }(\theta)$ is the transverse momentum at the given offset angle $\theta$, and $p_{\bot}^{(m)}(\theta)$ its peak value, $A_0=\epsilon E_0/(\omega\sqrt{1+\epsilon^2})$. Note that Eq.~(\ref{pbotmaxtheta}) is not equivalent to the similar Eq.~(\ref{pbotmax}), because of the momentum dependence of the $\theta$-$\phi_i$ relationship of Eq.~(\ref{attoclock-angle}), i.e., for the given attoclock angle $\theta$,  the corresponding $\phi_i$ is different for $p_{\bot}^{(m)}$ and $p_{\bot }$. Note also that the definition of $\alpha$ is different from the one in Ref.~\cite{Hartung_2021}, where $\alpha=(p_{\bot }-A_0)/(p_kA_0/c)$. In the Simpleman estimation it is $\alpha =1$.

In Fig.~\ref{fig3} we show the time-resolved  $\alpha$-parameter dependence on the ellipticity of the laser field. First of all, there is no significant effect of sub-barrier CC as TCSFA results coincide with the plain SFA at any ellipticity. The Simpleman results mostly coincide with the plain SFA besides large offset angles and small ellipticity, which indicates in the latter cases the role of the nonadiabatic corrections beyond the leading $\gamma$ terms. Note also that $\alpha=1$ for the Simpleman at $\delta \theta=0$, but $\alpha$ increasing at large offset angles. Significant deviation of CCSFA results from the plain SFA is observed in nonadiabatic regime at small ellipticity and large positive offset angles. This is due to nonadiabatic CC in the continuum. There is an asymmetry in $\alpha$ with respect to the sign of the offset angle due to nonadiabaticity.

In Fig.~\ref{fig4} we discuss the field dependence of the $\alpha$-parameter. As expected, for the plain SFA $\alpha=1$ at any ellipticity and intensity. There is a remarkable influence of the continuum CC, which  increases significantly the $\alpha$-parameter in weak fields and at small ellipticities. The Simpleman model does not fully account CC, especially nonadiabatic CC in the continuum.


\section{Longitudinal momentum at small ellipticity values: Role of recollisions}


In this section we check the capability of the improved Simpleman model for  small ellipticity values. In this case recollisions play a role and we have no analytical expression to account for the recollision effect. For this reason, we calculate the final longitudinal momentum of the electron numerically within classical consideration for the most probable trajectory, assuming that the initial momentum components of the electron at $\phi_i=0$ are
 \begin{eqnarray}
 p_{x i} &=&0, \\
 p_{y i}&=&\frac{\epsilon\gamma(\phi_i)  \kappa}{6}, \\
 p_{ki}&=&\frac{I_p}{3c},
\end{eqnarray}
and the initial coordinate of the tunnel exit is
 \begin{eqnarray}
  \mathbf{r}_e=-\frac{\kappa^2}{2E(\phi_i)^2}\mathbf{E}(\phi_i).
\end{eqnarray}
With these initial conditions  Newton equations are integrated assuming the Coulomb field as a perturbation:
 \begin{eqnarray}
 p_{k}=p_{ki}-\frac{\textbf{p}_{\bot_i }\cdot\textbf{A}(\phi_i)-A(\phi_i)^2/2}{c}-\int_{\eta_i}^\infty d\eta \partial_z V(\mathbf{r}(\eta,\eta_i)),\nonumber\\
\end{eqnarray}
with $\omega\eta_i=\phi_i$

In Fig.~\ref{fig5} we show the ellipticity dependence of the maximum of the longitudinal momentum, calculated  via the Simpleman estimation. Surprisingly, the given Simpleman estimation fits well to the experimental results of  Ref.~\cite{Maurer_2018}.

\section{Experimental time-resolved spectra vs CCSFA}

In this section we compare the CCSFA calculations of the time-resolved nondipole momentum shift  to the experimental data of Ref.~\cite{Willenberg_2019}. In the latter, rather than  the light-front momentum, the data for the longitudinal momentum  are presented.

In Fig.~\ref{fig6}   we show the offset angle, expressed as a time delay, between the attoclock angle of the maximum yield and the minimum of the longitudinal momentum vs ellipticity. Both the Simpleman and CCSFA results are in accordance with the experimental data within the error bars. However, the experimental data hints for a slight slope decreasing the time delay at large fields. This feature is absent in the Simpleman model, but demonstrated by  CCSFA.

In Fig.~\ref{fig7} the  CCSFA and Simpleman results for the average of the longitudinal momentum vs the attoclock offset angle are compared with the experiment. The Simpleman results on the time-resolved data coincide with that of CCSFA, however, there are significant deviations with respect to the experiment, especially at large offset angles and at large ellipticities.
This indicates that there is an unaccounted large Coulomb effect (larger at larger ellipticity) in the nondadiabatic regime (larger at small fields, i.e., at larger offset angles), or some time delay, which is very intriguing, and still unexplained.

In Fig.~\ref{fig8} we compare the results of CCSFA for the dependence of the average of the longitudinal momentum on the transverse one with the experimental data of Ref.~\cite{Hartung_2019}. Generally, the Coulomb corrections are not very significant for the given interaction regime. However, we note an important message of Fig.~\ref{fig8}, that the sub-barrier Coulomb corrections increase the momentum shift along the propagation direction $\langle p_k\rangle$, while the continuum one oppositely decreasing it, which is in accordance with Ref.~\cite{He_2021}.

\section{Conclusion}

We have developed a nondipole CCSFA theory and a improved Simpleman model, which include Coulomb corrections during the sub-barrier dynamics and in the continuum  up to  first order in $E_0/E_a$, the improved Simpleman model includes nonadiabatic corrections up to first order in $\gamma$. Both CCSFA and Simpleman model are applied for the description of the time-resolved (attoclock angle resolved) nondipole longitudinal dynamics. Further,  we have introduced the light-front momentum, which absorbs the trivial relativistic correlation between the transverse and longitudinal momenta and allows to elucidate the role of nonadiabatic and Coulomb effects. Our conclusion is that in the quasistatic regime the plain SFA and the Simpleman model describe quite well the time-resolved nondipole longitudinal dynamics, because of a partial compensation of the sub-barrier and the continuum Coulomb effects. In contrast, the nonadiabatic Coulomb effects, especially large at small ellipticity values, introduce a deviation of the Simpleman model and the plain SFA with respect to the full CCSFA. In particular,  the nonadiabatic Coulomb effect in the continuum violate the symmetry of the light-front momentum with respect to the sign of the attoclock offset angle. The Coulomb effect is especially conspicuous at small ellipticity $\epsilon \lesssim 0.6$ and positive offset angles, and  gives rise to interest for experimental observation, see for instance Figs.~\ref{fig1}(d), and \ref{fig3}(d). The same kind of CC induces a large deviation of the parameter $\alpha$ from the Simpleman value 1, see for instance the weak field region in Fig.~\ref{fig4}(a). The parameter $\alpha$ describes the shift of the peak of the transverse momentum distribution with respect to variation of the longitudinal momentum.

We find  deviations of CCSFA results from the experimental data of Ref.~\cite{Willenberg_2019} for large offset angles and large ellipticities, which indicate that  there is a notable nonadiabatic Coulomb effect and/or ionization time delay still remaining  not reproducible within our CCSFA based on the eikonal approximation and applicable only for soft rescatterings.

\acknowledgments

We thank Pei-Lun He for useful discussions, and Reinhard D\"orner and Benjamin Willenberg for providing experimental data.

\appendix

\section{Transverse nonadiabatic Coulomb momentum transfer}\label{A1}

Here we provide an intuitive estimation of the CC in the  direction transverse to the laser electric field in the polarization plane in the nondipole regime. Due to nonadiabaticity the electron obtains a transverse momentum during tunneling:
\begin{eqnarray}
   \delta p_{y\,i} =  \frac{\epsilon \gamma \kappa}{6} ,
\end{eqnarray}
where the $y$-axis is transverse to the field in the polarization plane.
The Coulomb momentum transfer can be estimated as
\begin{eqnarray}
   \delta p_{y\,C} &\sim& \frac{Z}{x^2}\frac{y}{x}\delta t,
\end{eqnarray}
where $x$ is the coordinate along the laser electric field direction, $y$ is the transverse displacement, and $\delta t$ is the effective interaction time with the atomic core. As $x \sim  E_0\delta t^2/2$, the effective time can be estimated as
\begin{eqnarray}
  \delta t\sim \sqrt{\frac{x_0}{E_0}},
\end{eqnarray}
assuming during this time the electron displacement is twice  the distance of the tunnel exit $x_0\sim \frac{I_p}{E_0}$.
The transverse  displacement is
  \begin{eqnarray}
 y &\sim & p_{yi} \delta t+ \int^{\delta t} dt'\int^{t''} dt'' E_y(t'') = p_{y i} \delta t +\frac{\omega E_0 \delta t^3}{6},
   \end{eqnarray}
where $E_y\sim \epsilon E_0\omega t$ is the transverse nonadiabatic force.  The first terms is estimated as
\begin{equation}
\label{CCC1}
   \sim \frac{\epsilon\gamma \kappa}{6}\frac{2 Z E_0}{\kappa E_a},
\end{equation}
and the second one as
\begin{equation}
\label{CCC2}
  \sim E_0\frac{\omega \delta t^3}{6}\sim  \frac{\epsilon\gamma \kappa}{6}\frac{2 Z E_0}{\kappa E_a}.
\end{equation}
Thus,
  \begin{eqnarray}
    \delta p_{y \,C } \sim  \frac{\epsilon\gamma \kappa}{6}\frac{4 Z}{\kappa}\frac{E_0}{E_a}.
\end{eqnarray}

\section{Longitudinal Coulomb momentum transfer}\label{A2}

Here we provide an intuitive estimation of the CC in the laser propagation direction in the nondipole regime, when the electron has an initial momentum at the tunnel exit $p_{ki}$. The Coulomb momentum transfer can be estimated as
\begin{eqnarray}
   \delta p_{k\,C} &\sim& \frac{Z}{x^2}\frac{z}{x}\delta t.
\end{eqnarray}
We estimate the longitudinal displacement:
  \begin{eqnarray}
  z &\sim & p_{ki} \delta t+\int \frac{p_\bot (t')}{c} E(t')dt'\sim p_{zi} \delta t+\frac{E_0^2\delta t^3}{6c}.
   \end{eqnarray}
   Thus
   \begin{eqnarray}
   \delta p_{k C} &\sim& \frac{Z}{x_0^3}p_{zi} \delta t+    \frac{Z}{x_0^3}\frac{E_0^2\delta t^3}{6c}.
\end{eqnarray}
The first terms is estimated as
\begin{equation}
\label{CCC1}
   \sim p_{ki}\frac{2 Z E_0}{\kappa E_a},
\end{equation}
and the second one as
\begin{equation}
\label{CCC2}
  \sim \frac{I_p}{3c}\frac{2 Z E_0}{\kappa E_a}.
\end{equation}
Taking into account that $p_{ki}=\frac{I_p}{3c}$, we have
  \begin{eqnarray}
    \delta p_{k\, C } \sim p_{ki}\frac{4 Z}{\kappa}\frac{E_0}{E_a}.
\end{eqnarray}

\section{Transverse and longitudinal momentum correlation}\label{A3}

Here we provide an alternative derivation of the coefficient $\alpha$ related to the correlation of the transverse and longitudinal components momentum. We solve the electron equations of motion in the laser field of magnetic dipole approximation, with a time dependent electric and magnetic field:
\begin{eqnarray}
  x''(t) &=& \frac{A_x'(t)}{\Lambda}, \\
  y''(t) &=&  \frac{A_y'(t)}{\Lambda},\\
  z''(t) &=& \frac{x'(t) A_y'(t)}{c\Lambda} - \frac{y'(t)A_x'[t] }{c\Lambda} ,
\end{eqnarray}
with the initial conditions $x'(t_i)=v_{xi}$, $y'(t_i)=v_{yi}$, $z'(t_i)=p_{zi}$, and $x(t_i)=y(t_i)=z(t_i)=0$.
The correction to the electron final momentum due to the electric quadrupole correction to the laser field is calculated perturbatively:
\begin{eqnarray}
   \Delta p_x &=& -\frac{1}{c}\int A_x(s) z(s) ds = -\frac{E_0 }{c\omega}p_{zi} \sin(\omega t_i)   \\
   \Delta p_y &=& -\frac{1}{c}\int A_y(s) z(s) ds = \frac{\epsilon E_0 }{c\omega}p_{zi} \cos (\omega t_i) .
\end{eqnarray}
The final transverse momentum is:
\begin{eqnarray}
   p_\bot (p_{zi})=\sqrt{v_{xi} - A_x(t_i) + \Delta p_x)^2 + (v_{yi} - A_y(t_i) + \Delta p_y)^2},\nonumber \\
\end{eqnarray}
 which we expand over the initial longitudinal momentum $p_{zi}$ around $p_{z0}$:
 \begin{eqnarray}
  \alpha &\equiv& \frac{p_\bot (p_{zi})}{\epsilon E_0/c\omega}\\
  &=& \left\{ \epsilon \cos(\omega t_i) \left[c v_{yi} \omega + \epsilon E_0 (c + p_{z0}) \cos (\omega t_i)\right]  \right. \nonumber\\
  &-&\left. c v_{xi}\omega\sin(\omega t_i)+ E_0 (c + p_{z0}) \sin^2(\omega t_i)\right\}\frac{1}{\epsilon}\nonumber\\
  &\times& \left\{c^2 ( v_{xi}^2 +  v_{yi}^2) \omega^2 + E_0 (c + p_{z0})\left[\epsilon\cos(\omega t_i) \left(2 c v_{yi}\omega \right.\right.\right. \nonumber  \\
  &+ & \left.\left.\left.  \epsilon E_0 (c + p_{z0}) \cos(\omega t_i)\right)   - 2 c v_{xi} \omega \sin(\omega t_i) \right.\right. \nonumber  \\
  & +&  \left.\left. E_0 (c + p_{z0}) \sin^2(\omega t_i)\right]\right\}^{-1/2}.\nonumber
  \end{eqnarray}
For the values for $v_{xi}$, and $v_{yi}$, we use Eqs.~(\ref{pbotCgi}), and for $ p_{z0}$, Eq.~(\ref{pkCgi}). The results for $\alpha$ with this estimation coincides with those in Figs.~\ref{fig3}.

\bibliography{strong_fields_bibliography}

\end{document}